# Towards a New and Compact Gas-Dynamic Cooler- Buncher for the FAIR Laspec and MATS Experiments

**Victor Varentsov**

*Facility for Antiproton and Ion Research in Europe (FAIR). Planckstraße 1. 64291 Darmstadt. Germany.*
  *E-mail*: `Victor.Varentsov@fair-center.eu`

ABSTRACT: In this article, we propose a new gas-dynamic ion beam cooler-buncher for the FAIR Laspec and MATS experiments. This very compact and having a simple design gas-dynamic cooler-buncher has a length of just 10.5 cm. It can convert continuous ion beams of various masses and with a transmission efficiency of around 100% into pulsed, high-quality ion beams. The special design of the supersonic nozzle and the RF-only funnel enables sufficient buffer gas (helium) thickness for effective ion beam cooling with an injection energy of up to several keV, without compromising the high vacuum in the other parts of the setup. The beams with a longitudinal emittance of down to 10 eV ns will be available. This emittance value is 20 times superior to the best current RFQ cooler-bunchers can deliver. We explored the operation of the proposed new gas-dynamic cooler-buncher by conducting detailed gas-dynamic and Monte Carlo ion-trajectory simulations. The results of these simulations are presented and discussed.

# Contents

## 1. Introduction

In the preceding quarter-century, radiofrequency quadrupole (RFQ) cooler-buncher devices have evolved into a pivotal component of the equipment utilized by numerous research centers engaged in high-precision nuclear and atomic physics experiments employing low-energy ion beams of both radioactive and stable elements. These RFQ devices perform the conversion of continuous ion beams into high-quality short bunches, characterized by low longitudinal and transverse emittance.

The construction of one of the first RFQ cooler-buncher units was completed in 1999 at the Ion Guide Isotope Separator On-Line (IGISOL) facility at the University of Jyväskylä in Finland [1, 2]. A review of the development of the IGISOL technique can be found in [3].

Descriptions of various RFQ cooler-buncher modifications and designs that are in operation, under commissioning or in development can be found in the review [4], as well as in dozens of original papers (e.g. [5-18]).

The new design of the RFQ cooler-buncher has recently been developed and commissioned off-line at the University of Jyväskylä. The authors call it HIBISCUS (Helium-Inflated Beam Improvement Setup that Cools and Undermines Spreads) and it is intended to be Finland's in-kind contribution to the MATS (Precision Measurements of very short-lived nuclei using an Advanced Trapping System for highly charged ions) and Laspec (Laser Spectroscopy of short-lived nuclei) experiments at FAIR (Facility for Antiproton and Ion Research facility). In April 2025, they published a high-quality paper [19] containing a detailed technical description of the characteristics, operating conditions and performance efficiency of this RFQ cooler-buncher.

Figure 1 shows the HIBISCUS experimental setup.

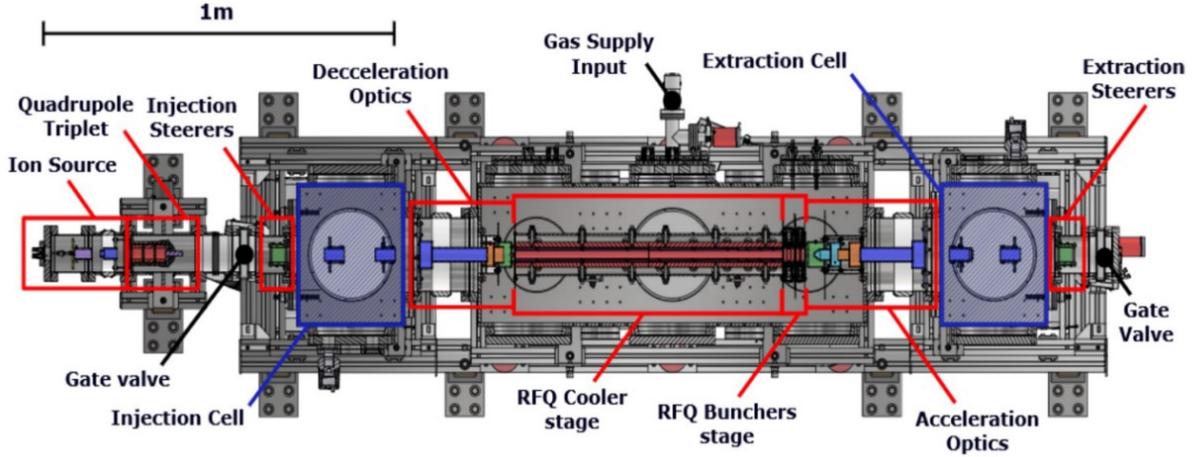

**Figure 1.** Top overview of the HIBISCUS beamline. The different parts of the experimental setup is identified. The injection and extraction cells, marked with hatched blue boxes, are designed to house beam diagnostics. The main central vacuum chamber, its supporting aluminium frame, pumping system and gas line are at high-voltage (HV). It is enclosed with an HV protection cage for safe operation, and to isolate it from the ground. The sub-assemblies, discussed in more detailed in Sec. 2 of the work [19], are marked with red boxes. Reproduced from [19]. CC By 4.0.

In 2020, we have proposed the concept of the windowless gas-dynamic ion beam cooler-buncher [20], which may provide fast and highly efficient deceleration, cooling and bunching of ion beams in a wide range of masses and injection energies up to 300 keV. This approach is somewhat alternative to all RFQ cooler-bunches described in the literature. It uses electromagnetic RF funnels and sufficiently dense supersonic and subsonic flows of buffer gas to slow down, cool and rapidly transport the injected ions to a vacuum. To become more familiar with this RF funnel technique, we recommend taking a look at our recent review [21], which also contains a description of the work [20], as well as our recent proposal [22] of a double-nozzle technique for in-gas-jet laser resonance ionisation spectroscopy.

It is quite clear that the design and operating conditions of the cooler-buncher [20], which are applicable for ions with injected energy up to 300 keV, cannot be considered as optimal for the ions with injected energy 10 times less (e.g. it is the case of the high precision Laspec experiments at FAIR). That is why we have developed a new compact gas-dynamic ion beam cooler-buncher for the FAIR Laspec and MATS experiments.

We have investigated this new version of the gas-dynamic cooler-buncher using detailed gas-dynamic and ion-trajectory Monte Carlo simulations. The results of these simulations are presented and discussed in the next sections. Similar simulations have been performed in [20] and are described elsewhere (e.g. see references [21-26]). Detailed gas dynamic simulations of the buffer gas flow we have made using the VARJET code. This code is based on the solution of a full system of time dependent Navier–Stokes equations and is described in detail in [27]. The results of the gas-dynamic simulations (flow fields of the buffer gas velocity, density and temperature) were then used in a Monte Carlo ion-trajectory simulations, where the electric fields were simulated using SIMION 8.1 [28].

## 2. General description and gas dynamics simulations

The new gas-dynamic ion beam cooler-buncher consists of three components, listed below:

1. The diverging supersonic nozzle with an inner conical tube on the axis for direct injection of the primary ion beam into the expanded supersonic buffer gas jet.

2. The RF-only funnel located on the axis between the exit plane of the nozzle and the RF butcher.

3. The cylindrical RF buncher consisting of 25 thin electrodes connected to the RF-only funnel by a metal diaphragm of 0.3 mm thickness and 1 mm aperture.

The assembled cooler-butcher's components are set on the 30 keV high-voltage platform, because the both injection beamline and the beamline to the Laspec equipment are at the ground potential. This is similar to the HIBISCUS setup [19].

Figure 2 and Figure 3 show the design of the proposed gas-dynamic cooler-buncher, along with the results of a detailed gas dynamic simulation for the helium velocity flow field (in the differentially pumped vacuum chamber of the RF-only funnel) and the flow field of gas pressure in the region of the RF buncher. A complex barrel shock wave structure of the gas flow inside RF-only funnel is clear visible in Figure 1.

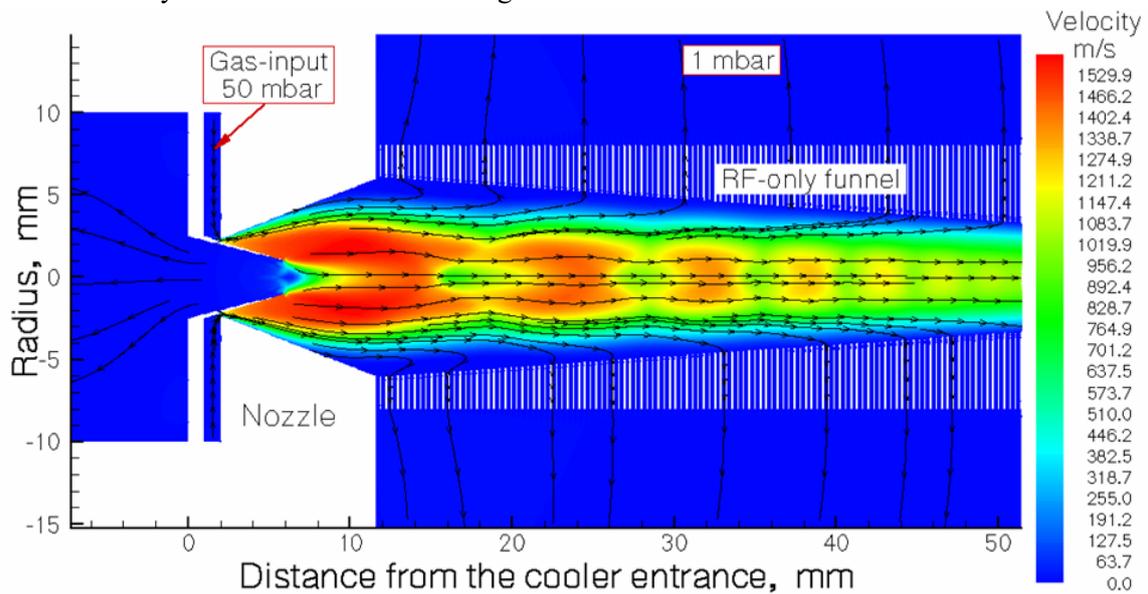

**Figure 2.** Schematic view of the nozzle (it has a central conical tube for the primary ion beam injection) and the entrance part of the FR-only funnel inside the cooler-buncher combined with the results of the gas-dynamic simulation for the helium velocity flow field. The stagnation input gas pressure and temperature are 50 mbar and 300 K, respectively. Black arrowed lines indicate the direction of gas flow. Details of the nozzle and RF-only funnel geometries are given in the text.

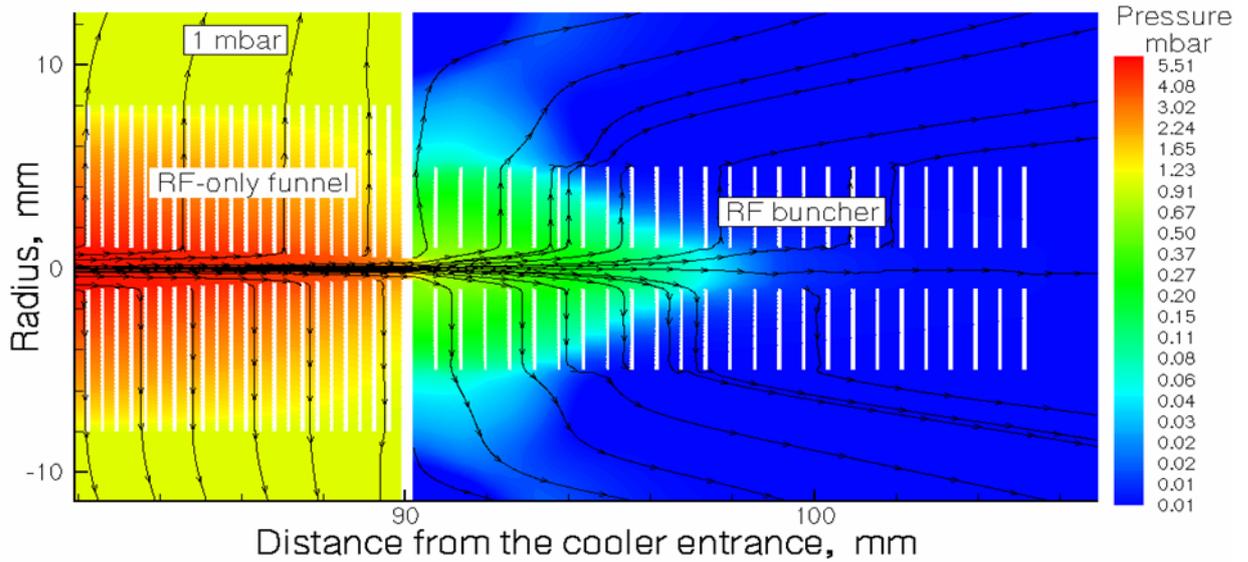

**Figure 3.** Schematic view of the end-part of the FR-only funnel and RF buncher combined with the results of the gas-dynamic simulation for the helium pressure flow field. The stagnation input gas pressure and temperature are 50 mbar and 300 K, respectively. Black arrowed lines indicate the direction of gas flow. Details of the nozzle and RF-only funnel geometries are given in the text.

The Figure 4 illustrates the data of Figure 3 for the helium pressure distribution along the axis.

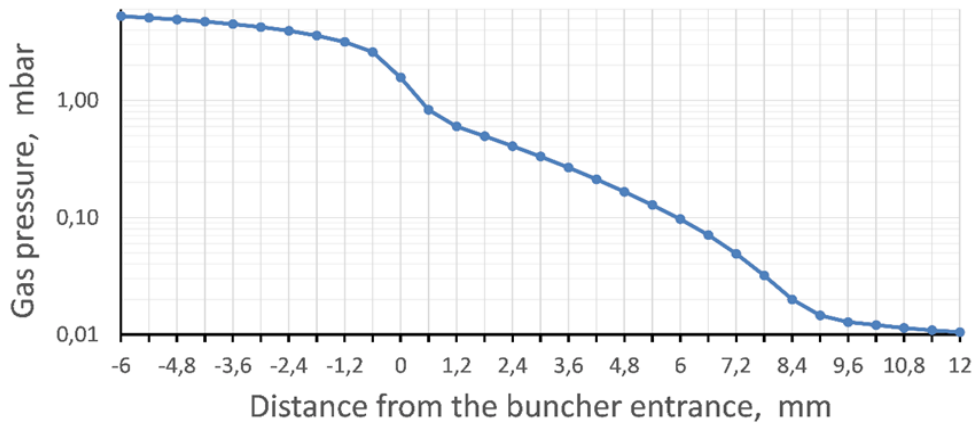

**Figure 4.** The calculated the helium pressure distribution along the axis. It illustrate the results of gas-dynamic simulation shown in Figure 3.

    The decreasing of the gas pressure (and density, of course) inside the cylindrical RF buncher (see it in Figure 4) do not allow for efficient transport the ions. Therefore, small DC electric field gradient is additionally applied along the buncher electrodes to drag the ions through this buncher and finally extract them into high vacuum conditions.
    The main design parameters of the supersonic nozzle, RF-only funnel and RF buncher are listed in Table 1 and Table 2. It should be noted that the total length of the cooler-buncher is only 10.5 cm (see Figure 3).

**Table 1**. The main design parameters of the nozzle.

| **Inner nozzle tube** |
| --- |

|  |  |
|---|---|
| Exit diameter | 2.0 mm |
| Entrance diameter | 5.0 mm |
| Total length | 6.0 mm |
| Supersonic part length | 4.0 mm |
| **Diverging nozzle** | |
| Throat diameter | 4.6 mm |
| Diverging cone length | 9.6 mm |
| Exit diameter | 12.0 mm |
| Annual gap between nozzle throat and inner tube in the nozzle | 0.1 mm |

**Table 2.** The main design parameters of the RF-only funnel and RF-buncher.

| **Design parameter** | **RF-only funnel** | **RF-buncher** |
|---|---|---|
| Entrance aperture diameter | 12.0 mm | 2.0 mm |
| Exit aperture diameter | 1.0 mm | 2.0 mm |
| Electrode thickness | 0.1 mm | 0.1 mm |
| Inter-electrode spacing | 0.25 mm | 0.5 mm |
| Number of electrodes | 220 | 25 |

Note that the design parameters of the nozzle, the RF-only funnel and the RF buncher, listed in Table 1 and Table 2 above, are very similar to those presented in our previous paper [20]. However, the design of the new device has undergone a few important changes, dictated by the much lower energies of the injected ions and the correspondingly lower gas jet densities required to decelerate them inside the RF-only funnel.

Firstly, the exit diameter of the diverging nozzle is equal to 12 mm, compared to 16 mm in [20].

Secondly, because of decreasing the nozzle exit diameter, the number of FR-only funnel electrodes became of 220, that on the 80 electrodes less compared to the "1st funnel" in the work [20].

Thirdly, there is no gap (of 13.65 mm) between the nozzle exit plane and the RF-only funnel inlet. This means that the second RF-only funnel, as mentioned in [20], is not needed.

Fourthly, the thickness of the electrodes in the RF buncher is reduced down to 0.1 mm, which is two times less than it is in [20]. This makes the process of pumping buffer gas through the RF buncher electrodes about 16% more efficient.

Results of gas-dynamic calculations for the gas flow rates through the vacuum chambers of the cooler-buncher for different stagnation input gas pressures presented in Table 3.

**Table 3.** The calculated gas flow rates through the differentially pumped chambers of the cooler- buncher for different stagnation input gas pressures. The background gas pressure in the RF-only funnel chamber is 1.0 mbar for each stagnation pressure variant. The background pressures in the vacuum chambers are determined as the ratio of the gas flow rate to the effective pumping capacity of a given chamber.

| Stagnation input gas pressure | Gas flow rate [mbar l/s] | | | |
|---|---|---|---|---|
| | Total through the nozzle | Into entrance chamber | Into RF-only funnel chamber | Into RF buncher chamber |
| 25 mbar | 11.45 | $<1 \cdot 10^{-3}$ | 11.3 | 0.063 |
| 30 mbar | 14.55 | $<1 \cdot 10^{-3}$ | 14.46 | 0.083 |
| 40 mbar | 20.7 | $<1\cdot10^{-3}$ | 20.4 | 0.23 |
| 50 mbar | 26.9 | $<1\cdot10^{-3}$ | 26.3 | 0.62 |

It is important to note that we did not choose this pressure range (25-50) mbar for gas dynamics simulations by chance.

Firstly, the background pressure in the RF bancher chamber should not exceed $1\cdot10^{-3}$ mbar. For example, at a pumping speed of 2200 l/s, as in the HIBISCUS setup [20], the background pressure in the RF bancher chamber will be $3\cdot10^{-4}$ mbar at the input stagnation gas pressure of 50 mbar according to the data in Table 3.

Secondly, the input stagnation gas pressure should not be too low for fast and effective ion transport through the RF-only funnel.

Note that there is no direct proportionality between the total gas flow rate through the nozzle and the input stagnation gas pressure (see Table 3). This can be explained by the viscous gas interaction with the nozzle walls. The same is true for the gas flow rates into the RF buncher vacuum chamber: 0.063 mbar l/s at 25 mbar and 0.62 mbar l/s at 50 mbar (see Table 3). In this case it is explained by the changes to the structure of the supersonic gas jet inside the RF-only funnel.

In 2016, we proposed a new off-line laser ablation ion source for Laspec and MATS [29]. Later, a setup of this source was developed and commissioned at TU Darmstadt [30, 31]. Recently, this setup has been used in the collinear spectroscopy experiment [32] at TU Darmstadt.

The key elements of this laser ablation ion source are the RF-only funnel connected to the RF buncher, as shown in Figure 5.

We have presented the above data on the laser ablation ion source at the TU Darmstadt because the technical details of the RF-only funnel and RF buncher design in this setup, as well as the method of their assembly and installation, can be used to develop the setup of the gas-dynamic cooler-buncher proposed in this paper.

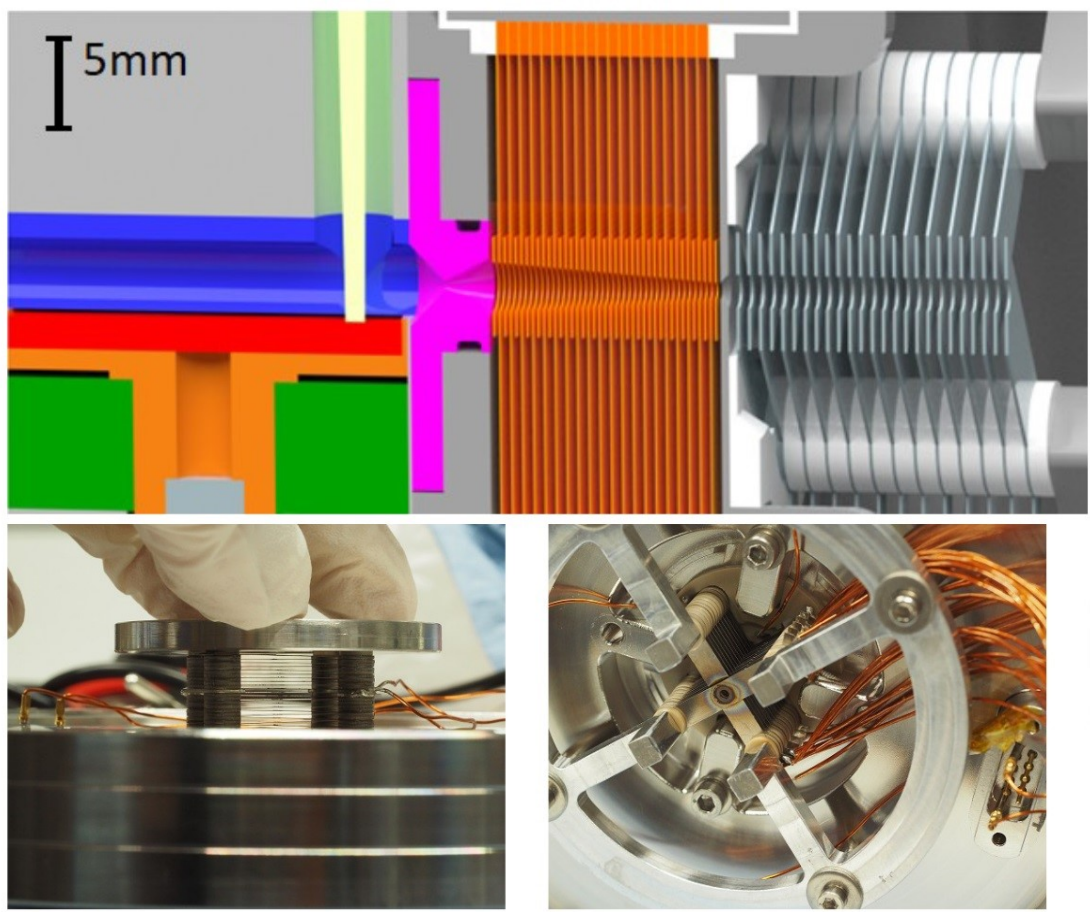

**Figure 5**. Laser ablation ion source setup at the TU-Darmstadt: Technical drawing of the ion source with ablation unit (color code as in the figure to the right): nozzle (magenta), RF-only funnel (orange), and RF buncher (gray) with insulators (white). Photos of the RF-only funnel (down left) and RF buncher (down right) which are ready for installation. Reproduced from [31].

## 3. Ion beam trajectory simulations

We ran detailed Monte Carlo ion-trajectory simulations were performed for both continuous and pulsed ion beam operation modes.

In these simulations, we used the following initial geometry for the injected beams:
- The distance from the entrance of the conical tube in the nozzle to the upstream point is 10 mm (see Figure 2).
- The beams have a diameter of 4 mm.
- The beams are focused into the nozzle with a focal length of 25 mm.

The RF frequency applied to both the RF-only funnel and the RF buncher is 10 MHz.

The RF amplitudes (in peak-to-peak, $V_{pp}$) applied to the RF-only funnel are 10 Vpp for ion masses up to 25-30, 15 $V_{pp}$ for ion masses up to 80, and 20 $V_{pp}$ for ion masses heavier than 80.

The RF amplitude applied to the RF buncher is 100 $V_{pp}$ for all ion masses.

– 7 –

It is important to note that the stability of ion movement inside the RF-only funnel and RF buncher does not depend on the mass-over-charge ratio. This is different from classic RF coolers, where the ratio does matter.

The calculated path of each ion extracted into a high vacuum ends at a distance of 8 mm from the exit plane of the RF buncher.

### 3.1 Transmission efficiency of ion beams of various masses and with different injection energies

The following four tables present the results of the Monte Carlo ion-trajectory simulations for the total transmission efficiency of ions of various masses and injection energies through the cooler-buncher. 'Injection energy' is defined as the ion energy relative to the electric potential of the nozzle.

**Table 4.** The cooler-buncher transmission efficiency in [%] for various ion masses (M) and injection ion beam energy (E). The stagnation input gas pressure is $P_o = 25$ **mbar** for all calculation variants. The number of extracted ions for each case is 2000 (it means ~4.5% statistical error for transmission efficiency of 100%).

| M  E (keV) | 25 | 50 | 75 | 100 | 125 | 150 | 175 | 200 | 225 | 250 |
|---|---|---|---|---|---|---|---|---|---|---|
| 0.05 | 89.1 | 94.1 | 95.7 | 91.5 | 96.8 | 96.9 | 96.4 | 93.6 | 94 | 66.4 |
| 0.5 | 85 | 93.7 | 92.2 | 90.5 | 96.4 | 93.8 | 91.6 | 89.1 | 89.5 | 56.7 |
| 1 | 82.1 | 89.8 | 88.2 | 85.3 | 89.3 | 84.2 | 82.4 | 79.5 | 71.8 | 42.1 |
| 2 | 79 | 80.7 | 75.5 | 70.3 | 75.5 | 65.1 | 65 | 49.4 | 29.2 | 22.7 |
| 4 | 57.9 | 63.3 | 54.7 | 47.5 | 44.4 | 39.9 | 30.6 | 25.1 | 12.4 | 7.5 |
| 8 | 44.4 | 36.4 | 28.9 | 20.4 | 17 | 10.9 | 6.6 | 3.6 | 1.3 | 0.65 |
| 16 | 23.9 | 14.8 | 7.9 | 3.9 | 2.3 | 1 | 1 | 0.4 | 0.5 | 0.4 |
| 24 | 13.6 | 6.7 | 2.7 | 1 | 0.4 | 0.4 | 0.4 |  | 0.4 |  |
| 32 | 9.2 | 3.3 | 1 |  |  |  |  |  |  |  |
| 40 | 6.7 | 1.67 | 0.4 |  |  |  |  |  |  |  |
| 48 | 4.3 | 1 |  |  |  |  |  |  |  |  |
| 56 | 2.8 |  |  |  |  |  |  |  |  |  |

**Table 5.** The cooler-buncher transmission efficiency in [%] for various ion masses (M) and injection ion beam energy (E). The stagnation input gas pressure is $P_0 = 30$ **mbar** for all calculation variants. The number of extracted ions for each case is 2000 (it means ~4.5% statistical error for transmission efficiency of 100%).

| M  E (keV) | 25 | 50 | 75 | 100 | 125 | 150 | 175 | 200 | 225 | 250 |
|---|---|---|---|---|---|---|---|---|---|---|
| 0.05 | 93.3 | 96.8 | 96.3 | 96.3 | 97.5 | 95.3 | 95.3 | 95.1 | 94.3 | 93.2 |
| 0.5 | 91.7 | 97.2 | 96.1 | 96.7 | 96.7 | 96.4 | 94.5 | 93.2 | 89.4 | 90 |
| 1 | 88.1 | 93.5 | 96.3 | 93.3 | 93 | 92.9 | 86.4 | 80.2 | 78.7 | 71 |
| 2 | 86.2 | 83.1 | 84.3 | 80.6 | 76.4 | 71 | 66.8 | 66.7 | 53.9 | 50.5 |
| 4 | 66.4 | 69 | 61.7 | 54.7 | 52.2 | 44.7 | 40.1 | 34.2 | 28.5 | 22.4 |

| E (keV) | 25 | 50 | 75 | 100 | 125 | 150 | 175 | 200 | 225 | 250 |
|---|---|---|---|---|---|---|---|---|---|---|
| 8 | 47 | 40.9 | 33.7 | 27.3 | 19.3 | 15.5 | 10.9 | 6.7 | 4 | 2.7 |
| 16 | 21.9 | 18.9 | 11.1 | 5.6 | 3 | 1.6 | 0.7 | 0.4 | 0.4 | 0.4 |
| 24 | 15.6 | 7.5 | 3.2 | 1.23 | 0.45 | 0.3 | 0.3 | | | |
| 32 | 10.2 | 4.1 | 1.5 | 0.4 | | | | | | |
| 40 | 6.3 | 2.1 | 0.5 | | | | | | | |
| 48 | 5.1 | 1.1 | 0,4 | | | | | | | |
| 56 | 3.2 | 0.75 | 0.3 | | | | | | | |

**Table 6.** The cooler-buncher transmission efficiency in [%] for various ion masess (M) and injection ion beam energy (E). The stagnation input gas pressure is $P_0 = 40$ **mbar** for all calculation variants. The number of extracted ions for each case is 2000 (it means ~4.5% statistical error for transmission efficiency of 100%).

| M \ E (keV) | 25 | 50 | 75 | 100 | 125 | 150 | 175 | 200 | 225 | 250 |
|---|---|---|---|---|---|---|---|---|---|---|
| 0.05 | 92 | 96.6 | 98.6 | 98 | 98.5 | 98.1 | 98 | 97.8 | 95.9 | 98.8 |
| 0.5 | 94.8 | 97.9 | 96.4 | 98,6 | 99 | 98.9 | 97.7 | 97.9 | 96.3 | 98.5 |
| 1 | 92 | 96.2 | 98.6 | 98.1 | 97 | 96.7 | 95.2 | 93.3 | 90.9 | 87.1 |
| 2 | 88.8 | 92.3 | 90 | 87.7 | 89.8 | 89 | 82.3 | 81.7 | 76.1 | 82.4 |
| 4 | 77 | 79 | 77.6 | 73.9 | 69.6 | 67.2 | 59.2 | 55.6 | 46.3 | 60.8 |
| 8 | 56.1 | 53.8 | 53.5 | 48.8 | 41.1 | 35.3 | 28.6 | 26.1 | 18.6 | 32.2 |
| 16 | 37.4 | 31 | 24.1 | 18.2 | 12.8 | 9.5 | 5.8 | 3.9 | 1.4 | 6.9 |
| 24 | 23.9 | 19.7 | 12.5 | 7.64 | 4.2 | 2.3 | 1 | 0.55 | 0.4 | 1.2 |
| 32 | 18.6 | 11.5 | 5.8 | 3.5 | 1.3 | 0.5 | 0.4 | 0.3 | | 0.3 |
| 40 | 12.3 | 7.6 | 3.2 | 1.1 | 0.47 | 0,3 | | | | |
| 48 | 10.4 | 4.7 | 1.8 | 0.7 | 0.4 | | | | | |
| 56 | 8 | 2.8 | 0.9 | 0.33 | | | | | | |

**Table 7.** The cooler-buncher transmission efficiency in [%] for various ion masses (M) and injection ion beam energy (E). The stagnation input gas pressure is $P_0 = 50$ **mbar** for all calculation variants. The number of extracted ions for each case is 2000 (it means ~4.5% statistical error for transmission efficiency of 100%).

| M \ E (keV) | 25 | 50 | 75 | 100 | 125 | 150 | 175 | 200 | 225 | 250 |
|---|---|---|---|---|---|---|---|---|---|---|
| 0.05 | 91 | 96.1 | 97.4 | 98.5 | 98.6 | 98.9 | 98.3 | 98.3 | 98.6 | 98.8 |
| 0.5 | 91.8 | 96.1 | 99.3 | 98.8 | 98.1 | 98.4 | 98.6 | 98.6 | 99.1 | 98.5 |
| 1 | 93.3 | 96.1 | 98.0 | 97.70 | 97.7 | 97.9 | 97.3 | 96.9 | 95.6 | 95.1 |
| 2 | 85.0 | 91.8 | 93.8 | 94.30 | 93.0 | 91.6 | 88.2 | 89.0 | 84.0 | 82.4 |
| 4 | 82.0 | 83.7 | 84.2 | 81.70 | 77.6 | 74.8 | 82.0 | 67.4 | 64.4 | 60.8 |
| 8 | 78.1 | 66.8 | 63.4 | 59.10 | 57.2 | 50.4 | 45.7 | 41.1 | 36.9 | 32.2 |
| 16 | 48.8 | 44.2 | 35.7 | 31.40 | 26.4 | 21.5 | 16.5 | 12.4 | 10.4 | 6.9 |
| 24 | 33.6 | 28.8 | 21.9 | 16.80 | 12.3 | 8.3 | 5.4 | 3.8 | 2.1 | 1.2 |
| 32 | 26.7 | 19.0 | 13.3 | 8.90 | 5.9 | 2.9 | 1.8 | 1.0 | 0.7 | 0.3 |

| 40 | 23.0 | 11.9 | 8.5 | 4.60 | 2.5 | 1.2 | 0.6 | 0.4 | 0.4 | |
| 48 | 16.1 | 10.1 | 5.3 | 2.7 | 1.3 | 0.6 | 0.4 | 0.4 | | |
| 56 | 13.9 | 7.5 | 3.4 | 1.4 | 0.7 | 0.4 | 0.4 | 0.4 | | |

It is important to note that losses of ion beams due to increased injection ion energy only occur inside the RF-only funnel. This is because the DC electric field gradient causes all ions to effectively travel through the RF buncher.

Figures 6 and 7 illustrate the data in Tables 4–7. Figure 6 shows the total transmission efficiency of 100-mass-unit ions through the cooler-buncher as a function of ion injection energy. Figure 7 shows the total transmission efficiency of ions of different masses.

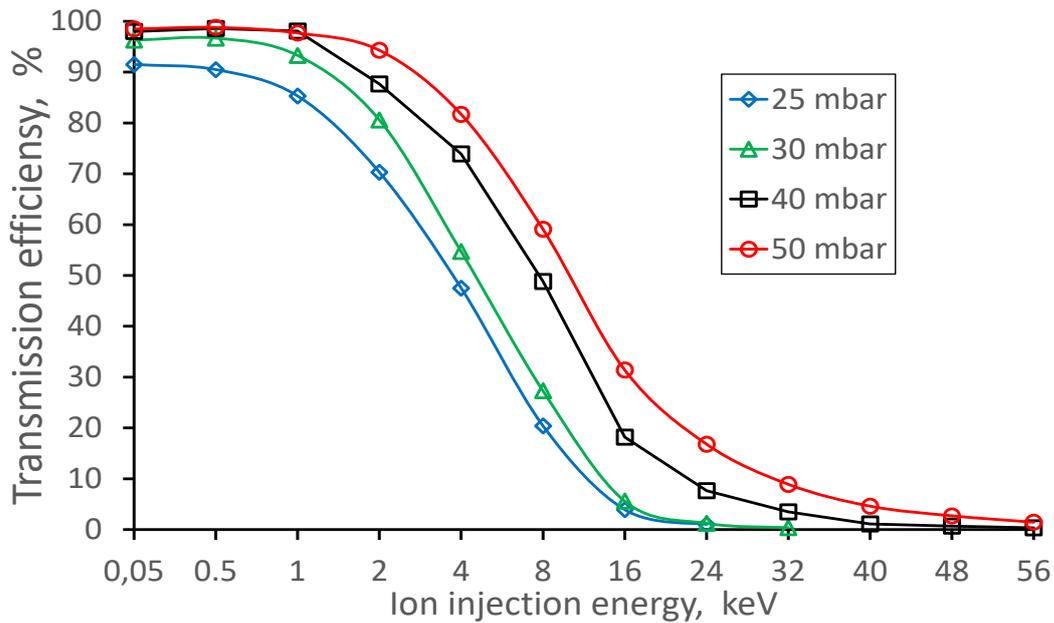

**Figure 6.** Results of the Monte Carlo ion-trajectory simulations for the total transmission efficiency through the cooler-buncher as a function of ion injection energy. Ion mass is M = 100. The number of extracted ions for each stagnation input gas pressure case is 2000.

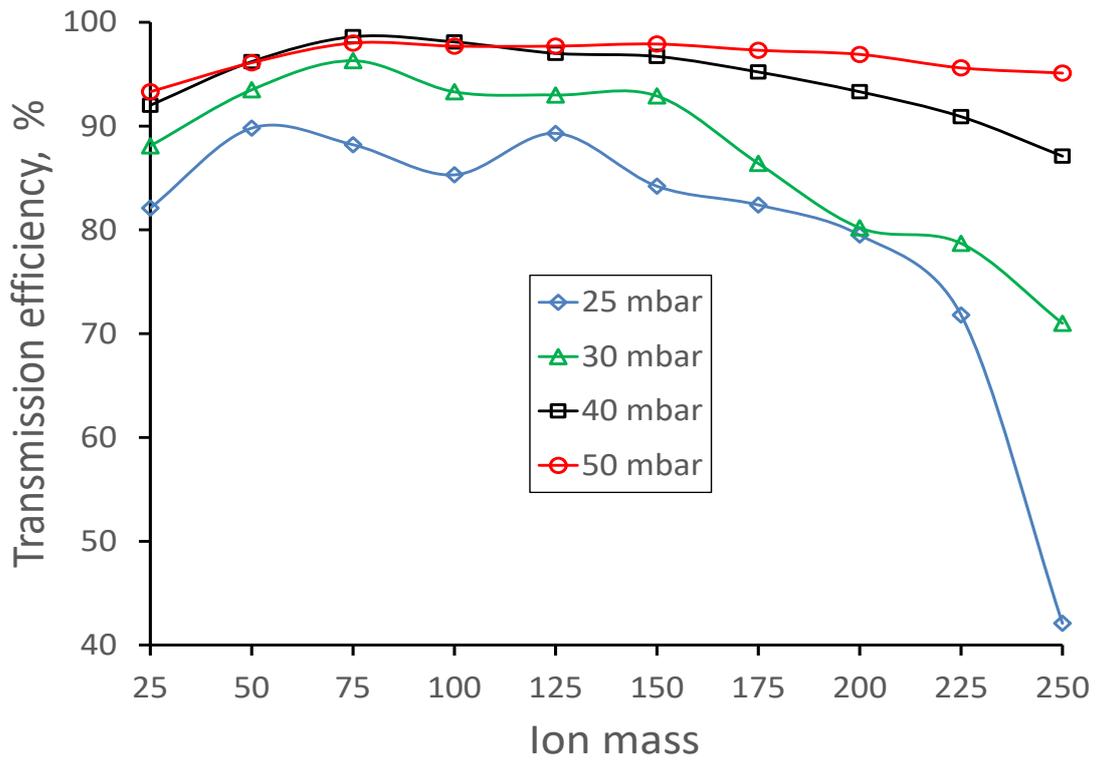

**Figure 7.** Results of the Monte Carlo ion-trajectory simulations for the total transmission efficiency through the cooler-buncher as a function of ion mass. Ion injection energy is equal to 1 keV. The number of extracted ions for each stagnation input gas pressure case is 2000.

Figure 8 shows the results of the Monte Carlo calculations for the time-of-flight distribution of 100-mass-unit ions through the cooler-buncher in continuous-beam operating mode at various gas stagnation pressures.

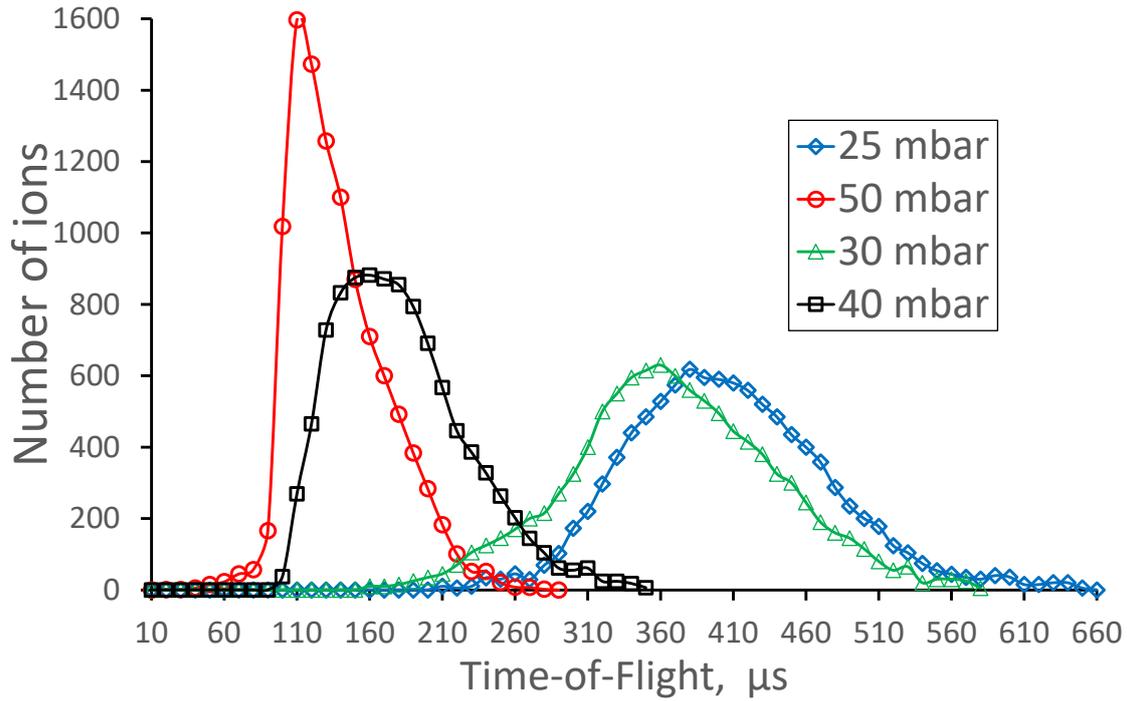

**Figure 8.** These are the results of the Monte Carlo ion-trajectory simulations for the time-of-flight distribution of 100-mass-unit ions through the cooler-buncher in continuous-beam operation mode at different input gas stagnation pressures. The ion injection energy is 1 keV. The DC electric field gradient is 0.5 V/mm. The number of ions extracted for each case is 10,000.

The different shapes and positions of the time-of-flight spectrums shown in Figure 8 are explained by changes in the complex gas flow structure inside the RF-only funnel at different gas stagnation pressures.

Table 8 shows the results of the Monte Carlo simulations for the characteristics of the extracted ion beams with a mass of 100 in continuous-beam operation mode at different stagnation input gas pressures.

**Table 8**. Calculated characteristics of the extracted ion beams with a mass 100 in continuous-beam operation mode at different stagnation input gas pressures. Ion injection energy is equal to 1 keV. The DC electric field gradient is 1.0 V/mm. The acronym FWHM signifies Full-Width at Half-Maximum.

| Stagnation input gas pressure (mbar) | 25 | 30 | 40 | 50 |
|---|---|---|---|---|
| Longitudinal velocity (m/c) | 6388 | 6362 | 6312 | 6250 |
| energy (eV) | 20.4 | 20.23 | 19.91 | 19.52 |
| Longitudinal (FWHM) velocity spread (m/c) | 227 | 227 | 288 | 263 |
| energy spread (eV) | 0.0537 | 0.0537 | 0.0864 | 0.072 |
| Radial velocity (m/s) | 150 | 175 | 200 | 200 |
| energy (eV) | 0.0234 | 0.0319 | 0.0208 | 0.0208 |
| Radial (FWHM) velocity spread (m/s) | 229 | 258 | 253 | 286 |
| energy spread (eV) | 0.0273 | 0.0347 | 0.0333 | 0.0423 |
| Beam radius (90%) (mm) | 0.72 | 0.65 | 0.65 | 0.64 |
| Transverse emittance $\varepsilon_{x,y}$ ($\pi \cdot mm \cdot mrad$) | 11.83 | 12.61 | 14.56 | 14.48 |

| Normalized emittance $\varepsilon^N_{x,y} = \varepsilon_{x,y} \cdot [E]^{1/2}$ ($\pi \cdot$mm$\cdot$mrad $\cdot$[eV]$^{1/2}$) | 53.47 | 56.717 | 64.97 | 63.98 |
|---|---|---|---|---|

The data in Table 8 (for a stagnation input gas pressure of 50 mbar) show that the normalised transverse emittance is 64·π·mm·mrad·(eV)$^{1/2}$, corresponding to an emittance value of 0.37·π·mm·mrad upon final extraction in a high vacuum at an energy of 30 keV. The extracted continuous ion beam also exhibits very low longitudinal and radial energy spreads of 72 and 42 meV, respectively.

### 3.2 Ion bunching

To shape a potential well inside the RF buncher, which uses a DC electric field gradient to transport ions, a positive electric potential must be applied to one of the buncher's electrodes during ion trapping. We believe that the electrode on the RF buncher (see Figure 3) that is optimal for ion trapping is the 10$^{th}$ electrode, which is 6 mm from the entrance of the RF buncher (see Figure 4). This trapping electrode does not require an applied RF voltage. Only a fast-switched voltage is needed to trap the ions inside the RF buncher and extract them as a short pulse.

With a DC electric field gradient of 1.0 V/mm, the voltage difference between two adjacent buncher electrodes is 0.6 V. We found that applying +20 V to the trapping electrode is appropriate. For lower values of the DC electric field gradient, the trapping voltages should be proportionally lower.

The Monte Carlo simulations show that the ions of various masses squished into the bunch inside the potential wall in the region of the 7th electrode of the RF buncher (or at 4.2 mm distance from the buncher entrance – see Figure 4). The ion bunch is shaped like a cylinder. It is about 1 mm in diameter and about 0.5 mm long.

The results of the Monte Carlo simulations for the characteristics of extracted ion beams with a mass of 100 in pulsed-beam operation at various stagnation input gas pressures are presented in Tables 9 and 10 for DC electric field gradients of 0.5 V/mm and 1 V/mm, respectively.

**Table 9.** Calculated characteristics of the extracted ion beams with a mass 100 in pulsed-beam operation mode at different stagnation input gas pressure. Ion injection energy is equal to 1 keV. The DC electric field gradient is 0.5 V/mm.

| Stagnation input gas pressure (mbar) | 25 | 30 | 40 | 50 |
|---|---|---|---|---|
| Longitudinal velocity (m/c) | 4025 | 4025 | 4025 | 4025 |
| energy (eV) | 8.10 | 8.10 | 8.1 | 8.1 |
| Longitudinal (FWHM) velocity spread (m/c) | 175 | 172 | 200 | 173 |
| energy spread (eV) | 0.0319 | 0.0308 | 0.0417 | 0.0312 |
| Radial velocity (m/s) | 200 | 200 | 200 | 200 |
| energy (eV) | 0.0208 | 0.0208 | 0.0208 | 0.0208 |
| Radial (FWHM) velocity spread (m/s) | 425 | 440 | 430 | 415 |
| energy spread (eV) | 0.0941 | 0.1008 | 0.0963 | 0.0897 |
| Beam radius (90%) (mm) | 1.49 | 1.65 | 1.53 | 1.45 |
| Transverse emittance $\varepsilon_{x,y}$ ($\pi \cdot$mm$\cdot$mrad) | 52 | 58 | 53.76 | 50.94 |

| Normalized emittance $\varepsilon^N_{x,y} = \varepsilon_{x,y} \cdot [E]^{1/2}$ ($\pi \cdot mm \cdot mrad \cdot [eV]^{1/2}$) | 148 | 165.07 | 153 | 144.98 |
|---|---|---|---|---|
| Bunch time (FWHM) width (µs) | 0.96 | 0.84 | 0.85 | 0.53 |
| Longitudinal emittance (eV·ns) | 37.4 | 25.9 | 35.4 | 16.5 |

**Table 10**. Calculated characteristics of the extracted ion beams with a mass 100 in pulsed-beam operation mode at different stagnation input gas pressure. Ion injection energy is equal to 1 keV. The DC electric field gradient is 1 V/mm.

| Stagnation input gas pressure (mbar) | 25 | 30 | 40 | 50 |
|---|---|---|---|---|
| Longitudinal velocity (m/c) | 5725 | 5725 | 5725 | 5725 |
| energy (eV) | 16.38 | 16.38 | 16.38 | 16.38 |
| Longitudinal (FWHM) velocity spread (m/c) | 180 | 184 | 187 | 137 |
| energy spread (eV) | 0.0337 | 0.0353 | 0.0364 | 0.0196 |
| Radial velocity (m/s) | 285 | 300 | 300 | 275 |
| energy (eV) | 0.0422 | 0.0468 | 0.0468 | 0.0393 |
| Radial (FWHM) velocity spread (m/s) | 310 | 245 | 212 | 260 |
| energy spread (eV) | 0,05 | 0,0313 | 0,0234 | 0,0352 |
| Beam radius (90%) (mm) | 1.32 | 1.29 | 1.15 | 1.08 |
| Transverse emittance $\varepsilon_{x,y}$ ($\pi \cdot mm \cdot mrad$) | 46.47 | 47.8 | 42.18 | 36.68 |
| Normalized emittance $\varepsilon^N_{x,y} = \varepsilon_{x,y} \cdot [E]^{1/2}$ ($\pi \cdot mm \cdot mrad \cdot [eV]^{1/2}$) | 188.07 | 193.46 | 170.71 | 148.45 |
| Bunch time (FWHM) width (µs) | 0.50 | 0.52 | 0.48 | 0.5 |
| Longitudinal emittance (eV·ns) | 16.8 | 18.4 | 17.5 | 9.8 |

Results of the Monte Carlo simulations for the temporal distribution and the longitudinal velocity distribution in the extracted ion-pulse at different trapping times are presented in Figures 9 and 10, respectively. It is important to note that the width of the extracted ion bunch remains the same when the trapping time is decreased, even down to 0.125 ms (see Figure 9). You could say the same about the spread of the longitudinal velocity distribution shown in Figure 10. We assume in the Monte Carlo simulations that all ions spend the same amount of time during the bunching process. But this is not correct for a continuous ion beam because the ions entering the RF buncher first spend more time oscillating inside the potential well of the RF buncher than those who enter later.

Nevertheless, for trapping times of about 1–1.5 ms, the extracted ion beam characteristics listed in Tables 9 and 10 can be considered correct because more than 90% of ions have spent >0.125 ms in the bunching process.

Also, the distance between the 10[th] trapping electrode and the 7th electrode (where the bunch is located) is 1.8 mm. This means that ~2 µs is enough time to extract all of the buncher's ions from the trapping well in a DC electric field gradient of 1 V/mm. This means that the RF buncher can operate at a frequency of up to 1 kHz without losing injected ions. This is possible by periodically switching off the positive voltage applied to the 10th electrode for 2 µs.

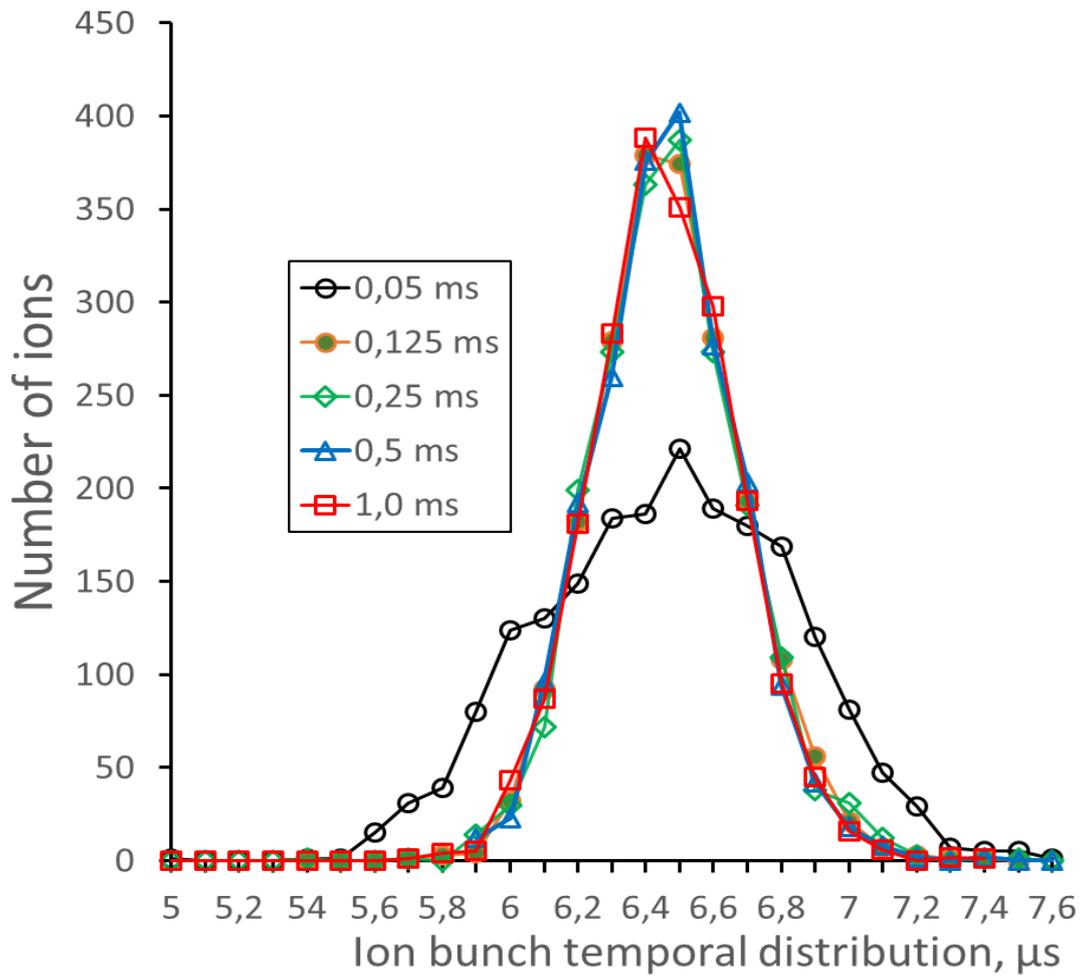

**Figure 9.** Results of the Monte Carlo simulations for the ion pulse temporal distribution at different trapping times. The ion mass is 100. The stagnation input gas pressure is 50 mbar. The ion's injection energy is 1 keV. The DC electric field gradient is 0.5 V/mm. For each case of the trapping time, the number of extracted ions is 2000.

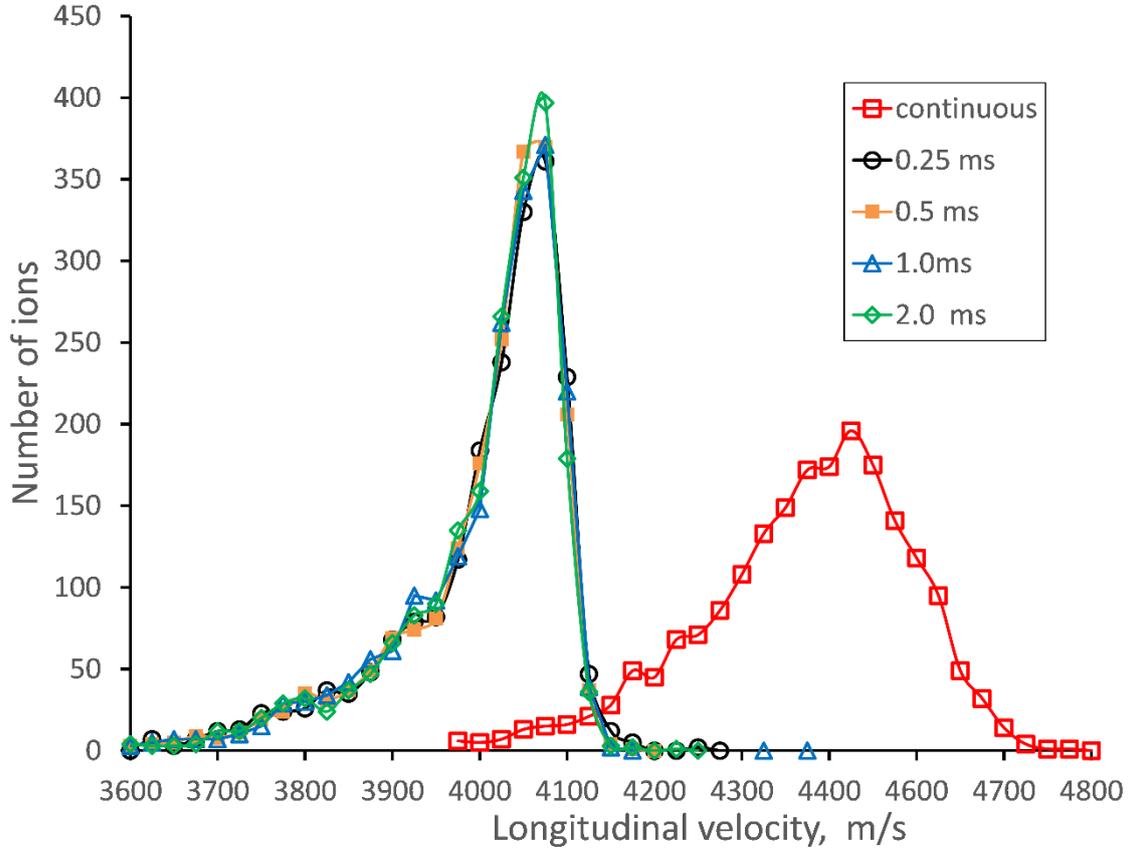

**Figure 10.** Results of the Monte Carlo simulations for the ion pulse longitudinal velocity distribution at different trapping times. The stagnation input gas pressure is 50 mbar. The ion mass is 100. The ion's injection energy is 1 keV. The DC electric field gradient is 0.5 V/mm. For each case of the trapping time, the number of extracted ions is 2000. The ion beam longitudinal velocity distribution in the continuous-beam operation mode is shown for comparison.

## 4. Discussion and outlook

The new compact gas-dynamic cooler-buncher presented in this paper and the classical RFQ cooler-buncher are in fact qualitatively different types of devices for their use in high-precision nuclear and atomic physics experiments with low-energy ion beams. This is so despite the fact that both types of devices use the RF field for radial ion's confinement during the ion beam cooling process.

In the RFQ cooler (e.g., see Figure 1), ion transport is achieved by an axially applied electric field gradient. In our gas-dynamic cooler, however, ions are transported through the RF-only funnel by supersonic and subsonic helium flow (see Figure 2).

It is well known that ion beam cooling is realized by a large number of collisions of ions with atoms of the buffer gas. The typical buffer gas pressure in the RFQ cooler chambers is in the range of 0.1–0.01 mbar (see, for example, references [17] and [19]). This gas can only be pumped through the inlet and outlet apertures. The length of the RFQ cooler chambers is therefore about 80–100 mm, in order to achieve the required aerial gas density for a proper ion beam termalization.

But since the pressure of the buffer gas inside the RF-only funnel, which is effectively pumped through the gaps between funnel electrodes (see Figures 2 and 3), is orders of magnitude greater, the length of the gas-dynamic cooler-buncher can be made much shorter (in the case presented here it is of 10.5 mm).

The next big advantage of our gas-dynamic cooler-buncher over all existing RFQ cooler-bunchers is that it can achieve transmission efficiencies of almost 100% for ion beams of various masses and injection energies up to 1 keV (see Tables 3–7 and Figures 6 and 7), and even higher. This is especially important for experiments involving radioactive ion beams, which have a very low production rate.

To ensure that the cooler-buncher operates properly, it is crucial to maintain a high vacuum in the adjacent vacuum chambers. This is why it is impossible to increase the buffer gas pressure inside the given RFQ cooler-buncher without a deterioration high vacuum in the system.

However, when using our gas-dynamic RF cooler-buncher, the supersonic buffer gas jet flowing around the conical inner tube of the nozzle closes behind its end (see Fig. 2)., This forms a strong rarefaction zone that prevents buffer gas from flowing into the high-vacuum chamber in front of the nozzle. The gas flow rate into the RF buncher vacuum chamber, which is open for pumping, is severely limited by the 1 mm diameter of the RF-only funnel outlet aperture, as well as the fact that over 98% of the total buffer gas flow rate through the nozzle is pumped through the gaps between the RF-only funnel electrodes. See Tables 2 and 3 for details.

A key finding of this paper is that, following a large number of computer experiments, we found that the proposed gas-dynamic cooler-buncher device would be capable of producing pulsed ion beams with world-record longitudinal emittances of up to 10eV ns. This value is 20 times lower than the lowest declared value for classic RFQ cooler-bunchers.

It is also worth noting that the new gas-dynamic cooler-buncher has an additional advantage over the classic RFQ bunchers. The buffer gas flow a few millimeters from the RF-only funnel exit (see Figure 3) helps to DC electric field gradient make the bunching process more efficient and faster by "pushing" ions towards the $10^{th}$ trapping electrode.

In conclusion, we can say that the new gas-dynamic cooler-buncher will be highly competitive with all existing RFQ cooler-bunchers once it is operational.

**Funding:** This research received no external funding.
**Data Availability Statement:** The data presented in this study are available upon request from the author.
**Conflicts of Interest:** The author declares no conflicts of interest